\documentclass[a4paper, oneside, twocolumn, notitlepage, 10pt]{extarticle_ecoc2015}
\usepackage{ecoc2015}
\usepackage{lipsum}
\usepackage{multicol}
\usepackage{ifthen}

\newcommand{\eig}[1]{\lambda_{#1}}      

\newcommand{\nfta}[1]{a(\lambda_{#1})}

\newcommand{\nftb}[2]{\ifthenelse{\equal{#2}{}}{b_{#1}(\lambda)}{b_{#1}(\lambda_#2)}}

\newcommand{\gbaud}{GBd}

\begin{document}
\selectlanguage{american}    


\title{Experimental Demonstration of Dual Polarization Nonlinear Frequency Division Multiplexed Optical Transmission System}%


\author{
    S. Gaiarin\textsuperscript{(1)}, A. M. Perego\textsuperscript{(2)},
    E. P. da Silva\textsuperscript{(1)}, F. Da Ros\textsuperscript{(1)},
    and D. Zibar\textsuperscript{(1)}
}

\maketitle                  


\begin{strip}
 \begin{author_descr}

   \textsuperscript{(1)} DTU Fotonik, Technical University of Denmark, Lyngby, 2800, Denmark
   \uline{simga@fotonik.dtu.dk}

   \textsuperscript{(2)} AIPT, Aston University, Birmingham B4 7E7 UK,
   \uline{peregoa@aston.ac.uk}

 \end{author_descr}
\end{strip}

\setstretch{1}


\begin{strip}
  \begin{ecoc_abstract}
    Multi-eigenvalues transmission with information encoded simultaneously in both orthogonal polarizations is experimentally demonstrated.
   Performance below the HD-FEC limit is demonstrated for 8-bits/symbol 1-\gbaud~ signals after transmission up to 207 km of SSMF.

  \end{ecoc_abstract}
\end{strip}


\section{Introduction}
The increasing demand for capacity and the limitations in achievable information rates imposed by fiber nonlinearities to the conventional coherent transmission systems have pushed part of the research community to investigate new techniques to encode and transmit information over the fiber channel.
Starting from the inverse scattering transform theory, the concept of nonlinear frequency division multiplexed (NFDM) system has been introduced few
years ago \cite{Mansoor}.
The first investigations focused on modulating only the position of a few discrete eigenvalues and transmitting over short distances.
Further extensive research enabled to demonstrate NFDM systems using multiple degrees of freedom (eigenvalue, spectral amplitude) transmitting through thousands of kilometers\cite{Turitsyn,son}.

Nevertheless, NFDM systems still suffer from many challenges
which have not yet allowed to match the spectral
efficiencies achievable by the technologically mature linear coherent transmission systems. One of the main limitations is the inability
of taking advantage of both polarizations supported by the standard single mode fiber (SMF), which can potentially double the spectral
efficiency of such systems. A first preliminary attempt in this direction has been reported by the numerical analysis of \cite{Maruta}.

In this work, exploiting a rigorous mathematical technique, we report, to the best of our knowledge, the first experimental demonstration of NFDM transmission independently modulating the two polarizations. The information
data is encoded in the phases of the four nonlinear Fourier (NF) coefficients $\nftb{}{}$ corresponding to two discrete eigenvalues, using
quadrature phase shift keying  (QPSK).
The system is able to reach a transmission distance of 207~km and 166~km at the hard-decision forward error correction (HD-FEC, BER=3.8$\times10^{-3}$) threshold, over a standard link  with
EDFA
amplification of
41.5~km and 83~km SMF fiber spans, respectively.
\begin{figure}[htp]
  \centering
   \includegraphics[width=\columnwidth]{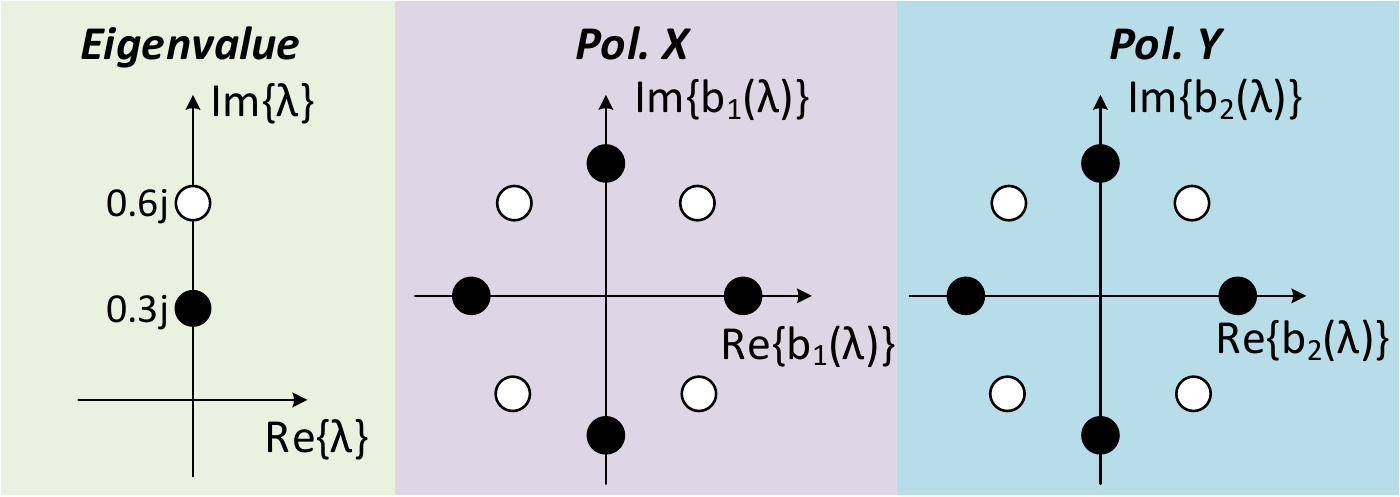}
  \caption{Ideal eigenvalues and corresponding spectral amplitude constellations diagrams.}
  \vspace{0.5cm}
  \label{fig:idealconst}
\end{figure}


\begin{figure*}[htbp]
  \centering
  \includegraphics[width=0.84\textwidth]{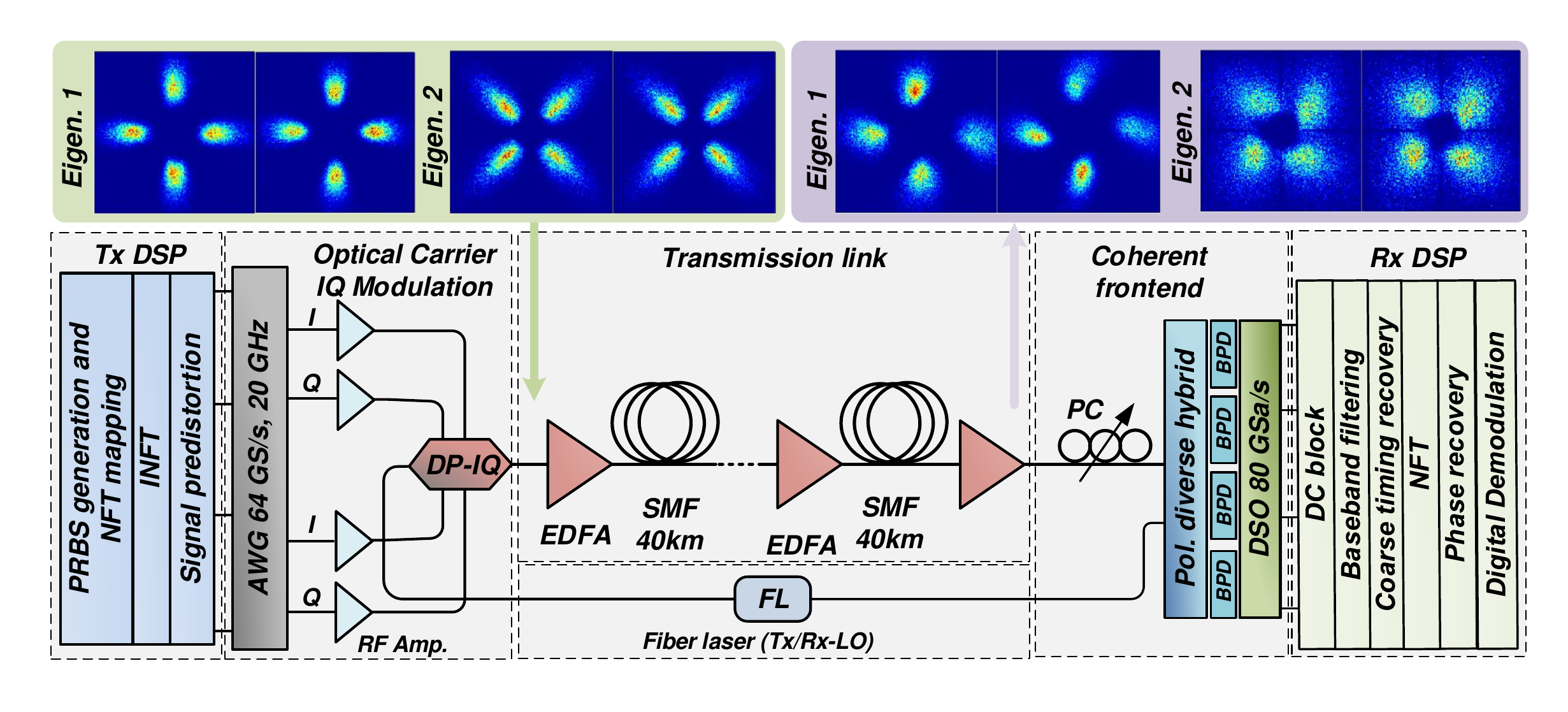}
  \caption{Experimental setup with transmitter and receiver DSP chain. Four constellations (two per polarization) associated to the two eigenvalues are shown at the transmitter side (top left) and after 373.5~km transmission with 41.5-km spans (top right).}
  \label{fig:setup}
\end{figure*}

\section{Mathematical framework}

The signal propagation in single mode fibers (SMFs) can be modeled by the so called Manakov system (MS) which describes the evolution
of the pair of complex field amplitudes $q_1$ and $q_2$ corresponding to the two polarizations of the field envelope .
The MS for the focusing regime in normalized form reads:
\begin{equation}
\begin{aligned}\label{eq:MS}
\frac{\partial q_1}{\partial z}&=&i\frac{\partial^2q_1}{\partial t^2}+i2q_1\left(|q_1|^2+|q_2|^2\right),\\
\frac{\partial q_2}{\partial z}&=&i\frac{\partial^2q_2}{\partial t^2}+i2q_2\left(|q_1|^2+|q_2|^2\right),
\end{aligned}
 \end{equation}
where $z$ and $t$ are the normalized space and time coordinates respectively.
The MS in Eqs.~\eqref{eq:MS} is integrable by the inverse scattering transform \cite{Manakov}, also called nonlinear Fourier transform (NFT).
It is therefore possible to associate a Lax pair of operators to the MS and define an eigenvalue problem with eigenvalues $\lambda$
and eigenvectors $\psi$. The time evolution differential equation for $\psi$ , induced by the eigenvalue problem, is
\begin{eqnarray}
\label{eq:eigenEvolution}
\partial_t\begin{pmatrix}
      \psi_1\\
      \psi_2 \\
     \psi_3
     \end{pmatrix}= \begin{pmatrix}
       -i \lambda& q_1 & q_2 \\[0.5em]
       -q_1^* & i\lambda  & 0 \\[0.5em]
       -q_2^* & 0 & i\lambda
     \end{pmatrix}\begin{pmatrix}
      \psi_1\\
      \psi_2 \\
     \psi_3
     \end{pmatrix}.
\end{eqnarray}
By integrating Eq.~\eqref{eq:eigenEvolution} using appropriate boundary conditions, the eigenvalues $\eig{}$ and
the NF coefficients $\{\nfta{}, \nftb{1}{}, \nftb{2}{}\}$ can be obtained. Compared to the single-polarization case $\{\nfta{}, \nftb{1}{}\}$), the MS has an additional
NF coefficient, $\nftb{2}{}$. This new degree of freedom can be used in a NFDM communication system to encode twice as many information bits.

The time domain waveform associated to a set of NF coefficients can be synthesized from a null field by using
the Darboux  transformation (DT) for the Manakov system.
Such a powerful technique allows to add iteratively new eigenvalues to
the NF discrete spectrum by updating the time domain signals correspondingly. The generic solution $\psi$ of \eqref{eq:eigenEvolution} is
transformed by the DT as \mbox{$\tilde{\psi}=(\lambda_0I_3-G_0)\psi$}, where $G_0=\Psi M_0\Psi^{-1}$,9
with
 \begin{eqnarray}
 \Psi= \begin{pmatrix}
       \psi_1 & \psi_2^* & \psi_3^* \\
       \psi_2 & -\psi_1^*  & 0 \\
       \psi_3 & 0 & -\psi_1^*
     \end{pmatrix}.
 \end{eqnarray}
The $\psi_j$ are the components of the eigenvalue problem solution that are being transformed, \mbox{$M_0=diag(\lambda_0,\lambda_0^*,\lambda_0^*)$}
and $\lambda_0$ is the new eigenvalue we want to add to the spectrum.
The signals are changed by the DT in the following way:
\begin{eqnarray}
 \tilde{q}_j=q_j+2i(\lambda_0^*-\lambda_0)\frac{u_j^*}{1+\sum_{k=1}^2|u_k|^2}, i=1,2
 \end{eqnarray}
being $\tilde{q}_j$ the modified component and \mbox{$u_j=\psi_{j+1}/\psi_1$}, where $j=1,2$ indicates one of the two polarizations.

%

\section{Experimental setup}
In order to demonstrate the advantage provided by the new degree of freedom $\nftb{2}{}$, the dual-polarization NFDM signaling depicted in Fig.~\ref{fig:idealconst} is implemented by choosing two reference eigenvalues $\eig{1}=0.3i$ and $\eig{2}=0.6i$.%

\begin{figure}[!hb]
  \includegraphics[width=0.95\columnwidth]{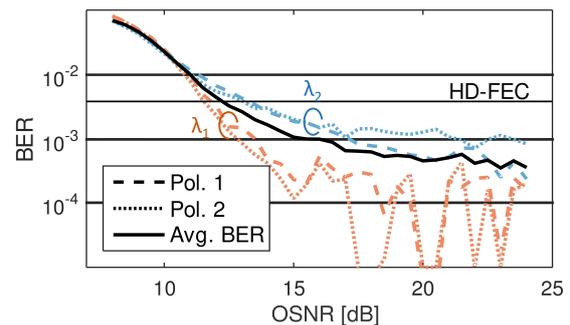}
  \caption{BER as a function of the OSNR in back-to-back.}
  \label{fig:b2bperfomance}
\end{figure}

\begin{figure*}[!htbp]
  \centering
  \includegraphics[width=0.337\textwidth]{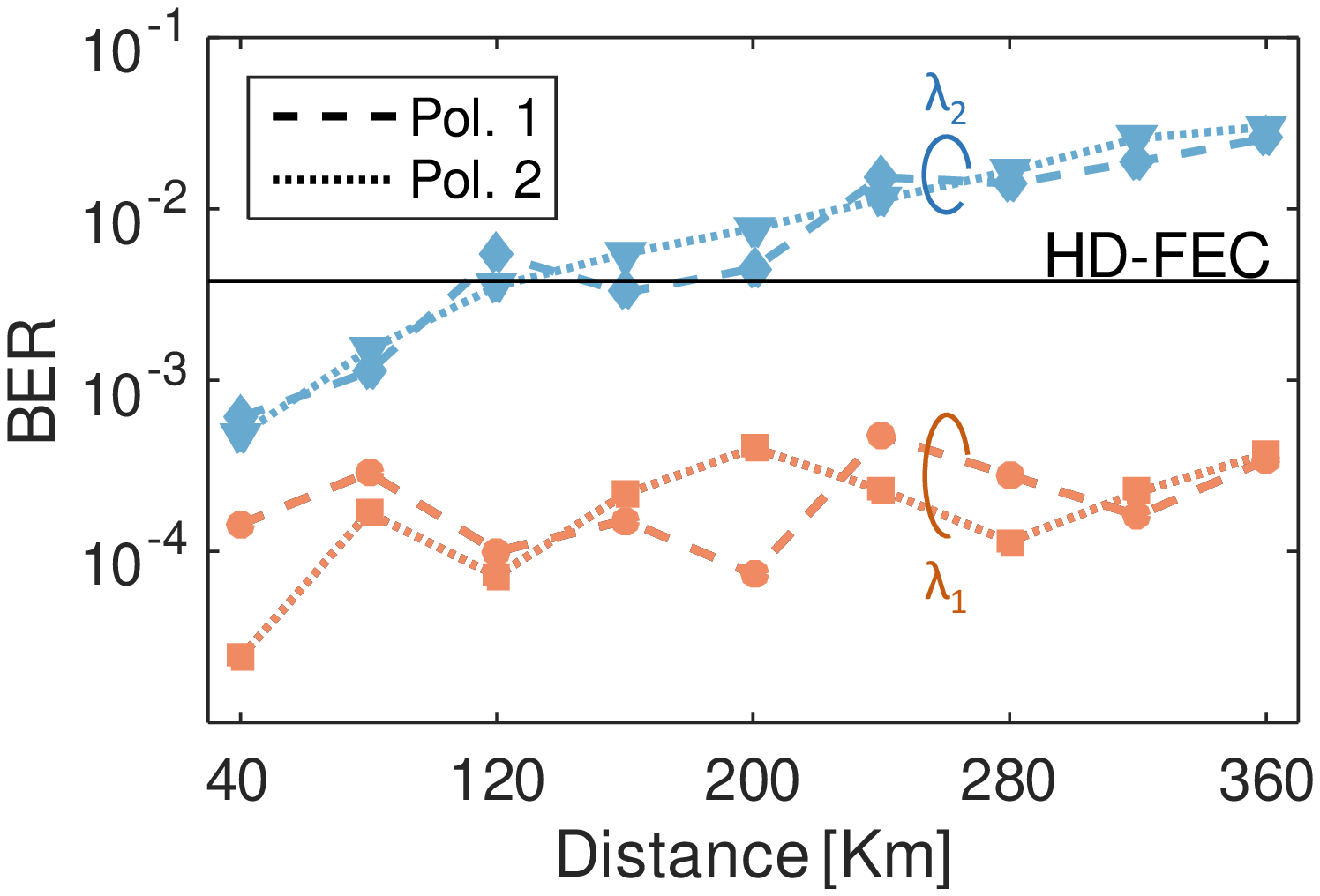}
  \includegraphics[width=0.295\textwidth]{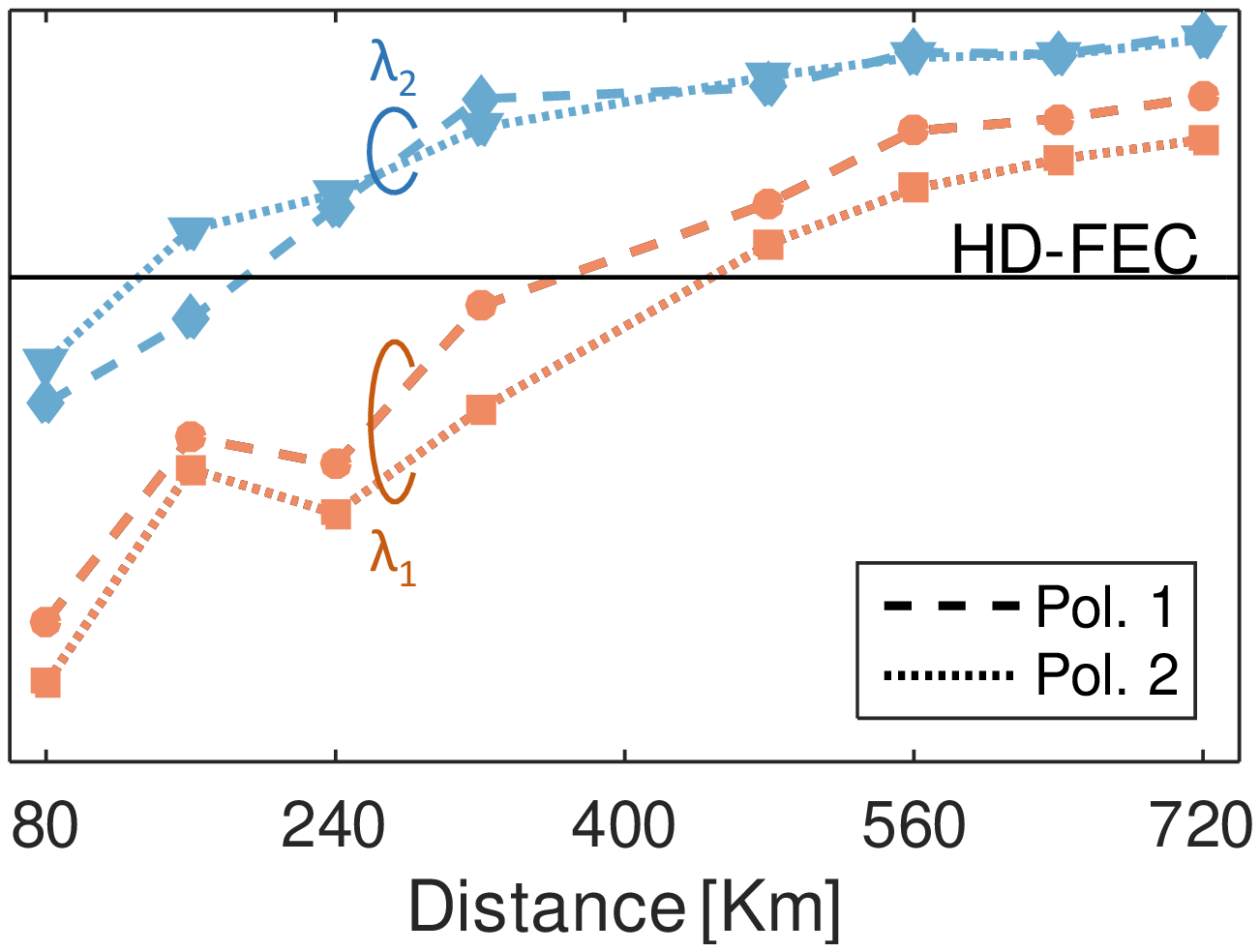}
  \includegraphics[width=0.295\textwidth]{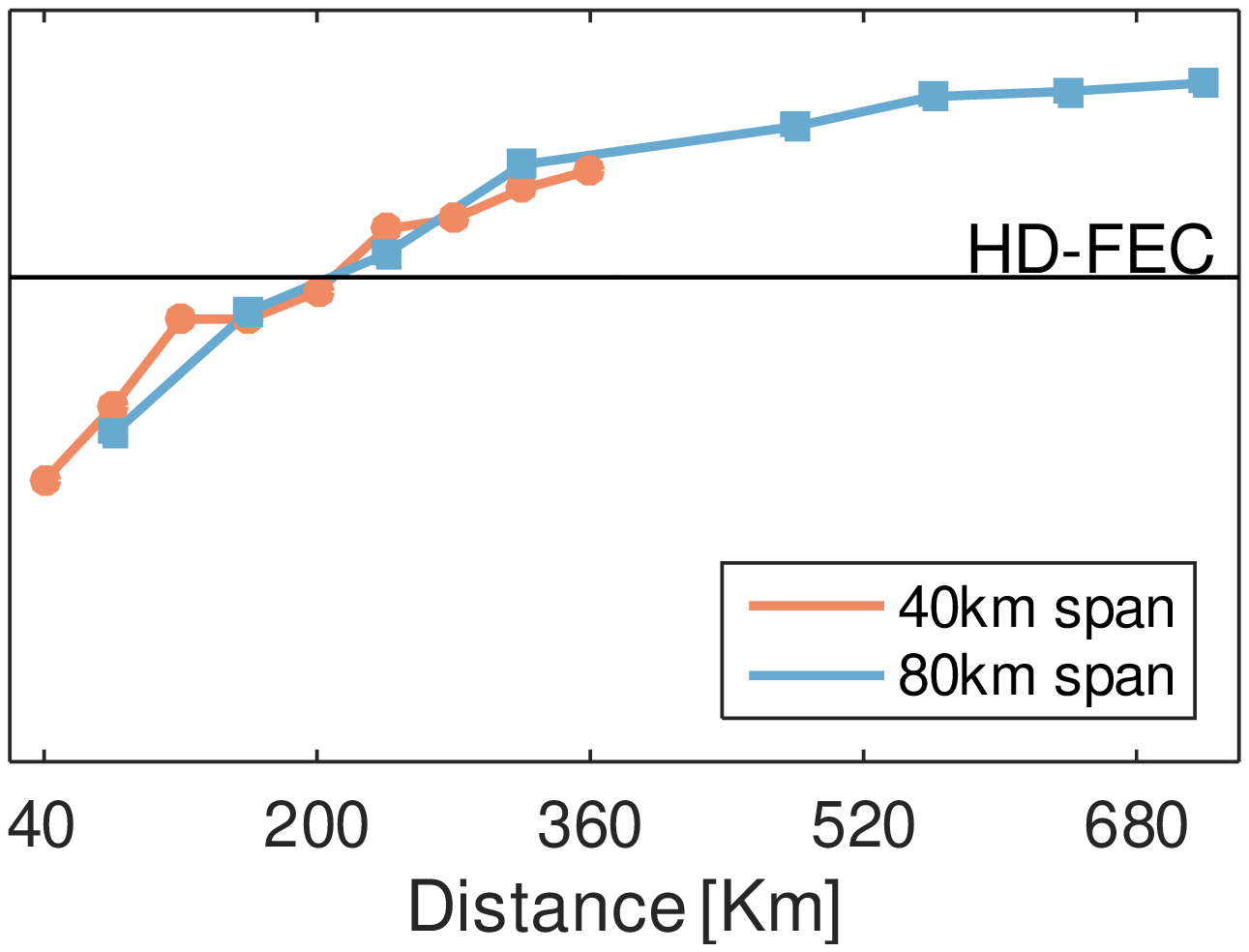}
  \caption{BER as a function of the transmission distance for each individual eigenvalue for 41.5~km (a) and 83~km (b) spans.
  (c)~Comparison of the average BER versus transmission distance
  between links of  the two different span lengths.}
  \label{fig:performance}
\end{figure*}
Data symbols are generated at 1~\gbaud~ by mapping the bits
from a pseudo random bit sequence of length  $2^{11}-1$  to the phases of each of the NFT coefficient pairs $\{\nftb{1}{i}, \nftb{2}{i}\}$
  according to a QPSK constellation for $\nftb{1,2}{1}$ and to a QPSK constellation rotated by $\pi/4$ for $\nftb{1,2}{2}$, (Fig~\ref{fig:idealconst}).

 The data so encoded are experimentally transmitted through the setup of Fig.~\ref{fig:setup}.
Inverse NFT (INFT) is used to generate time waveforms from the symbol $\{\nftb{1}{1}, \nftb{2}{1}, \nftb{1}{2}, \nftb{2}{2}\}$  using an iterative algorithm based on the DT (see previous section). The digital signal is pre-distorted to compensate for the Mach-Zehnder modulator nonlinear response and the waveforms are uploaded to an
 arbitrary waveform generator (AWG, 20-GHz analog bandwidth and 64-GSa/s) which drives a dual polarization I/Q
modulator. The optical input of the modulator is a fiber laser (FL) with $\sim$1~kHz linewidth, which is amplified after modulation and transmitted to a fiber link of multiple spans of SMF with dispersion $D$ =
17.5~ps/nm${\cdot}$km, nonlinear coefficient $\gamma$ = 1.25~W$^{-1}$km$^{-1}$ and attenuation $\alpha$ = 0.195~dB/km.

At the receiver, the polarization demultiplexing is performed manually through a polarization controller as no blind polarization demultiplexing algorithms for the NFT are yet available.
The signal is then detected in homodyne configuration by using a standard coherent receiver (33-GHz analog bandwidth, 80~GSa/s). The acquired
digital signal is rescaled to the ideal power of the signal generated by the INFT and low-pass filtered to remove the out of
band noise. FFT-based
coarse clock recovery is used in order to align the pulse to the center of the
processing time window. The signal is then passed to the NFT block which first locates the eigenvalues $\eig{i}$ (zeros of the NFT coefficient $\nfta{i}{}$) and then computes $\nftb{1,2}{i}$. In order to
minimize numerical errors, the NFT coefficients are computed by integrating \eqref{eq:eigenEvolution} using a generalized version of
the forward-backward  trapezoidal algorithm \cite{Aref}. The standard blind phase search algorithm is used independently on each of the constellations to perform carrier phase estimation.

\section{Experimental results}

The proposed system is first tested in a back-to-back (BTB) configuration by removing the transmission link
%
and varying the optical signal to noise ratio (OSNR) at the receiver input through noise loading. The measured signal bit
error rate (BER) is shown in Fig.~\ref{fig:b2bperfomance}. The BER averaged over the four constellations saturates toward a value of 5$\times 10^{-4}$ as the OSNR is increased. The main contribution of errors
comes from the two constellations of $\eig{2}$. As shown in Fig.~\ref{fig:setup} (top left) at maximum OSNR, both constellations related to $\eig{2}$ are more noisy and distorted at the transmitter due to the higher peak-to-average power ratio of the $\eig{2}$-pulse portion, and the limited digital to analog converter resolution. Additionally, imperfections in the digital signal processing (DSP) chain at the receiver can introduce further degradation. For example,  sub-optimum timing recovery due to noise, results in distortions of the received constellations, and an increment in  BER. Such a transceiver-induced penalty could be improved by more sophisticated DSP schemes.

The system is then tested by transmitting over a link of SMF fibers spans, first with spans of 41.5~km and then with spans of 83~km.  In
Fig.~\ref{fig:performance}~(a) and (b) the BER curves for the four different $\nftb{}{}$ are shown
as a function of the transmission distance for the 41.5~km and 83~km spans case respectively. As in Fig.~\ref{fig:b2bperfomance}, the BER is sensibly higher for
$\nftb{1,2}{2}$ as already hinted by the received constellations after a 373.5-km transmission (Fig.~\ref{fig:setup}). The constellations associated to $\eig{2}$ are more degraded while those associated to $\eig{1}$ are still well defined.
The performance of $\nftb{1}{i}$ and $\nftb{2}{i}$, which are associated to the two polarizations, are instead similar for the same eigenvalue $i$.
In Fig.~\ref{fig:performance}~(c)
we compare the average BER for the two span lengths. Theoretically, worse performance may be expected for the longer spans (further away from lossless transmission).
The results of Fig.~\ref{fig:performance}~(c) instead, do not show a relevant difference and the performance are similar in the two cases at the same transmission distance.
This is probably due to the  performance being mainly limited by transceivers impairments, so that
 the losses play a minor role.
Transmission distances up to 207 km for the configuration using 41.5~km long spans and up to 166~km for the one using spans of 83~km are achievable with a BER below
 the HD-FEC threshold of 3.8$\times$10$^{-3}$.

\vspace{-0.15cm}
\section{Conclusions}
The results presented in the paper demonstrate that polarization multiplexing in NFDM systems is feasible
and can lead to a potential increase of the spectral efficiency achieved up to now. The performance of the two polarizations fields reported are similar, while there is a clear difference in BER  between constellations associated to different eigenvalues.


\vspace{-0.15cm}
\section{Acknowledgements}
\begin{small}
This work is supported by the Marie Curie Actions through ICONE Project (no. 608099). We thank M. Kamalian
for the stimulating discussions.
\end{small}
\vspace{-0.2cm}
\bibliographystyle{abbrv}
\begin{spacing}{1.35}

\end{spacing}
\vspace{-4mm}

\end{document}